# Total transmission and total reflection of acoustic wave by zero index metamaterials loaded with general solid defects

Ziyu Wang,[1,2,3] Fan Yang,[3] LiBing Liu,[3] Ming Kang,[3] and Fengming Liu[1,a]

[1]School of Science, Hubei University of Technology, Wuhan 430068, China

[2]School of Physics and Technology, Wuhan University, Wuhan 430072, China

[3]Materials and Technology Institute, Dongfeng Motor, Wuhan 430056, China

[a] Corresponding author: fmliu@hbut.edu.cn

**Abstract:**

This work investigates acoustic wave transmission property through a zero index metamaterials (ZIM) waveguide embedded with a general solid defect. Total transmission and total reflection can be achieved by adjusting the parameters of the solid defect. We comprehensively study how longitudinal/transverse wave speeds of the solid defect affect the acoustic wave propagation through the waveguide. A two-dimensional (2D) acoustic crystals (ACs) system with zero index is used to realize these intriguing transmission properties. Thus, our work provides more possibilities to manipulate acoustic wave propagation through ZIM.

## I. INTRODUCTION

Zero index metamaterials (ZIM)，whose permittivity and permeability are simultaneously or individually near zero, have been studied both theoretically and experimentally and showed many intriguing properties.[1-10] Enoch et al.[2] showed that ZIM can be used to enhance the directive emission of an embedded source; Silveirinha and Engheta[3] presented an epsilon-near-zero (ENZ) medium that can "squeeze" electromagnetic (EM) wave through a very narrow channel, which has been demonstrated experimentally.[4,5] Hao et al.[6] first showed that total reflection or total transmission can be realized by introducing perfect electric (magnetic) conductor defects into the ZIM waveguide. Then, several works follow which concern manipulating EM wave propagation through ZIM waveguide by tailoring the parameters of the dielectric defect.[7-10] Meanwhile, acoustic metamaterials have also stirred considerable excitement for they can be engineered to exhibit intriguing physical phenomena.[11-22] Past efforts focused on realizing acoustic metamaterials exhibiting effectively negative bulk modulus and negative mass density.[11-14] Recently, acoustic ZIM has drawn intense attention from several research groups.[15-22] Among them, Qi et al.[22] introduced various kinds of defects in order to control acoustic wave transmission through the acoustic ZIM waveguide structure. However, they have only considered the liquid and ideal rigid defects. General solid defects, which are more complicated because of the co-existence of longitudinal and transverse waves but easier to realize in liquid host, still need further research

In this work, we investigate the acoustic wave transmission through the ZIM

waveguide embedded with general solid defects. Comprehensive analysis of how longitudinal/transverse wave speeds of the solid defect affect the transmission is provided, and numerical simulations are then carried out to testify our theory. In contrast to liquid defects, general solid defects provide more tunable parameters to control the transmission of acoustic wave through ZIM waveguide. In addition, a 2D ACs system, which can be mapped to an acoustic material with effectively zero density and zero reciprocal modulus [17], is used to realize these intriguing transmission properties.

## II. THEORETICAL ANALYSIS

A two-dimensional acoustic waveguide structure, which consists of four distinct regions, is illustrated in Fig.1. The regions 0 and 3 filled with water (with mass density $\rho_0$, bulk modulus $\kappa_0$ and speed $c_0 = \sqrt{\kappa_0/\rho_0}$) are separated by acoustic ZIM (region 1) with effective mass density $\rho_1$ and effective bulk modulus $\kappa_1$. And a cylindrical solid defect (region 2) with radius $a$, mass density $\rho_2$, bulk modulus $\kappa_2$, and shear modulus $\mu_2$ is embedded in the ZIM region. The walls of the waveguide are set to be hard walls.

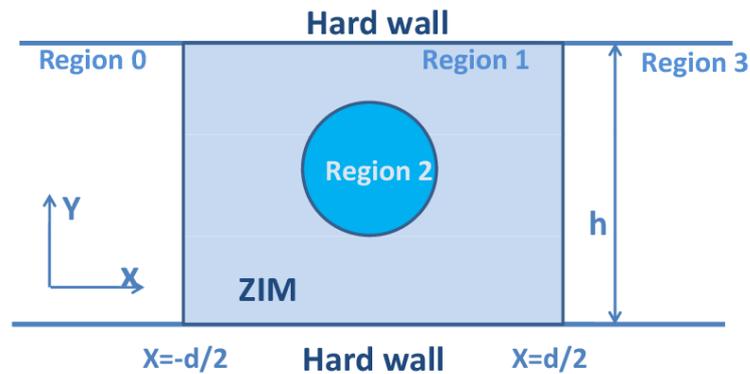

Figure 1. (Color online) Schematic of the 2D waveguide structure with water (region

0 and 3), ZIM (region 1), and an embedded defect (region 2).

Suppose that a plane harmonic acoustic wave $P_{inc} = Pe^{i(k_0 x - \omega t)}$ is incident from left to right inside the waveguide, where $P$ is the amplitude of the incident field, $k_0$ is the wave vector in the regions 0 and $\omega$ is the angular frequency. We omit the time variation item in the rest of this paper for convenience. Thus, the pressure and velocity fields in the region 0 can be written as

$$P_0 = P[e^{ik_0(x+d/2)} + Re^{-ik_0(x+d/2)}], v_0 = [e^{ik_0(x+d/2)} - Re^{-ik_0(x+d/2)}]P/\eta_0, \qquad (1)$$

while in region 3 the pressure and velocity fields must have the form

$$P_3 = TPe^{ik_0(x-d/2)}, v_3 = e^{ik_0(x-d/2)} PT/\eta_0, \qquad (2)$$

where $R$ and $T$ are the reflection and transmission coefficients and $\eta_0 = \sqrt{\rho_0 \kappa_0}$ is the impedance of the water host. In region 1, the pressure and velocity fields are presented by $P_1$ and $v_1 = (1/i\omega\rho_1)\nabla P_1$, respectively. In the limit $\rho_1 \approx 0$, the pressure field $P_1$ must be constant in the ZIM so as to keep $v_1$ as finite value. Then by using the continuous boundary condition at $x = -d/2$ and $x = d/2$, we have $P(1+R) = P_1$ and $TP = P_1$, which leads to $1 + R = T$. Applying the law of conservation of mass $\oint \rho v_r dl = \int (\rho/\kappa)(\partial p/\partial t) ds$ ($v_r$ denotes the normal component of the velocity field), we can find out the transmission coefficient of the waveguide as,

$$T = \frac{1}{1 - (\eta_0/2hp_1)\oint_{\partial A_2} v_{2r} dl}, \qquad (3)$$

where $\partial A_2$ is the boundary of the defect and $v_{2r}$ is the normal component of velocity field in the defect. For simplicity, only one defect is considered here. To evaluate $T$, we need to find the normal component of velocity field $v_{2r}$, which can be obtained by considering the elastic boundary conditions between the solid cylinder and the ZIM.

In region 2, the displacement fields are described by the elastic wave equation $(\lambda_2 + 2\mu_2)\nabla(\nabla \cdot \mathbf{u_2}) - \mu_2 \nabla \times \nabla \times \mathbf{u_2} + \rho_2 \omega^2 \mathbf{u_2} = 0$, where $\mathbf{u_2}$ is the displacement field, $\mu_2$ denotes shear modulus, and $\lambda_2$ represents the Lame constant satisfying $\kappa_2 = \lambda_2 + \mu_2$. We note that, in contrast to the liquid defects studied previously,[22] another elastic parameter $\mu_2$ needs to be considered here. In cylindrical coordinates, the general solution in the cylinders can be expressed as

$$\mathbf{u_2} = \sum_n [d_{n1}\mathbf{J}_{n1}(k_{2l}r) + d_{n2}\mathbf{J}_{n2}(k_{2t}r)], \tag{4}$$

where $\mathbf{J}_{n1}(k_{2l}r)$ and $\mathbf{J}_{n2}(k_{2t}r)$ are, respectively, defined as

$$\begin{aligned}\mathbf{J}_{n1}(k_{2l}r) &= \nabla[J_n(k_{2l}r)e^{in\phi}] \\ \mathbf{J}_{n2}(k_{2t}r) &= \nabla \times [\hat{\mathbf{z}}J_n(k_{2t}r)e^{in\phi}]\end{aligned}, \tag{5}$$

where $k_{2l} = \omega\sqrt{\rho_2/(\kappa_2 + \mu_2)}$, $k_{2t} = \omega\sqrt{\rho_2/\mu_2}$, $J_n(x)$ is the Bessel function and $n$ is the angular quantum number. The elastic boundary conditions require that

$$\begin{aligned}u_{1r}|_{r=a} &= u_{2r}|_{r=a} \\ p_1|_{r=a} &= \tau_{2rr}|_{r=a} \\ 0 &= \tau_{2r\phi}|_{r=a}\end{aligned}, \tag{6}$$

where $u_r$ is the normal component of displacement, $\tau_{rr}$ is the normal projection of stress tensor, and $\tau_{r\phi}$ is the tangential projection of stress tensor. We note that, to satisfy the elastic boundary conditions, the angular quantum number $n$ in Eq. (5) can only take 0 due to the fact that the pressure field $P_1$ is constant in the ZIM region. The elastic boundary conditions lead to the following equations

$$\begin{aligned}u_{1r} &= D_{11}d_{01} + D_{12}d_{02} \\ p_1 &= D_{21}d_{01} + D_{22}d_{02} \\ 0 &= D_{31}d_{01} + D_{32}d_{02}\end{aligned}, \tag{7}$$

where $D_{12} = D_{22} = D_{31} = 0$, $D_{11} = -k_{2l}J_1(k_{2l}a)$, $D_{32} = -k_{2t}^2 J_1(k_{2t}a)$, and $D_{21} = -2(\mu_2 + \lambda_2)k_{2l}J_1(k_{2l}a)/a + (\lambda_2 + 2\mu_2)k_{2l}^2 J_2(k_{2l}a)$ while $n$ takes 0. By solving Eq.

(7), we obtain $u_{1r} = u_{2r} = \frac{D_{11}}{D_{21}} p_1$. Then, as the velocity field is the derivative of the displacement field with respect to time, we have $v_{2r} = \frac{\omega}{i} u_{2r} = \frac{\omega}{i} \frac{D_{11}}{D_{21}} p_1$. Finally, the transmission coefficient (Eq. (3)) can be expressed as

$$T = \frac{1}{1 + \frac{\omega \eta_0 \pi a k_{2l} J_1(k_{2l}a)}{ihD_{21}}}. \qquad (8)$$

A deep inspection of Eq. (8) shows the transmission characteristics of the system. First, we can see that total transmission arises if $D_{11} = 0$, i.e. $J_1(k_{2l}a) = 0$. Although the divergence of $D_{21}$ can also lead to $T = 1$, tailoring $\mu_2$ cannot make $D_{21} = -2(\mu_2 + \lambda_2)k_{2l}J_1(k_{2l}a)/a + (\lambda_2 + 2\mu_2)k_{2l}^2 J_2(k_{2l}a)$ diverge obviously. Therefore, total transmission cannot be achieved by tailoring the transverse wave speed of the defect. We also would like to remark that as long as the longitudinal wave speed of the defect is fast enough, $J_1(k_{2l}a)$ will approach zero and thus total transmission happens, which is just like the case of ideal rigid defect.[21] Second, we will discuss total reflection. From Eq. (8), we see that if $J_1(k_{2l}a) \neq 0$ and $D_{21} = 0$, then $T = 0$, in which case total reflection happens. The expression of $D_{21}$ reveals that the transverse wave speed of the defect is involved and $J_1(k_{2l}a)$ and $J_2(k_{2l}a)$ should have the same sign to guarantee that $D_{21}$ has the possibility of equaling zero. Therefore, all the parameters of the solid defect contribute to achieving total reflection.

## III. NUMERICAL SIMULATION

To verify the analysis, numerical simulations are carried out by using the finite element method (FEM). We set $d = h = 2.7 \text{m}$. The wavelength $\lambda_0$ of the incident wave is $0.55$m. The ZIM with $\rho_1 = 0.0001\rho_0$ and $1/\kappa_1 = 0.0001(1/\kappa_0)$ is impedance match to water.

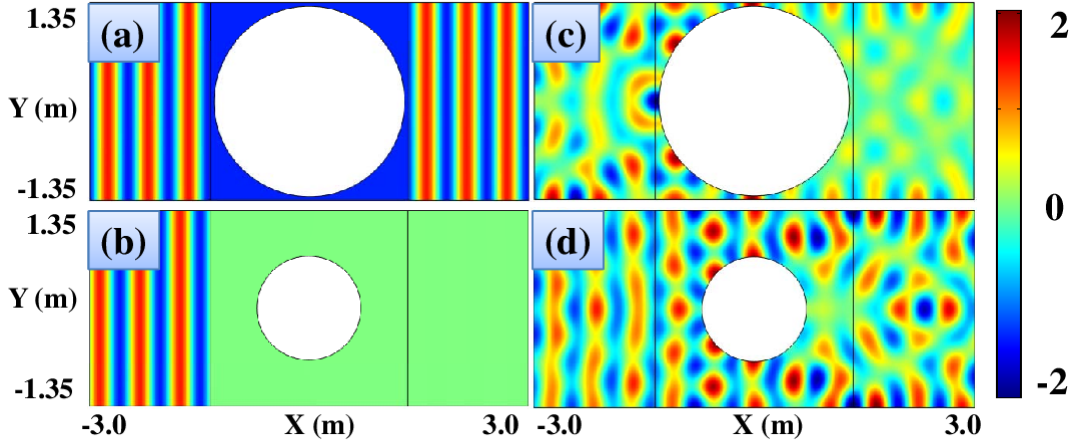

Figure 2. (Color online) The numerically simulated pressure field distributions for the ZIM waveguide structures realizing total transmission (a) and total reflection (b). The radii of the steel defects are $a = 1.3012$m and $a = 0.7121$m, respectively. (c) and (d) Comparative simulations for the waveguides without ZIM.

In Fig. 2(a), we show the pressure field distribution of the ZIM waveguides embedded with a steel ($\rho_2 = 7890$kg/m$^3$, $c_{2l} = 5780$m/s and $c_{2t} = 3220$m/s) defect to achieve total transmission. To satisfy $J_1(k_{2l}a) = 0$, we choose the radius of the defect $a = 1.3012$m. The plot clearly demonstrates that total transmission of the incident waves occurs even though the gap between the defect and the walls of the waveguide is very small. In Fig. 2(b), we show the pressure field distribution of the ZIM waveguides as total reflection happens. We consider a steel defect with radius $a = 0.7121$m, which is chosen to satisfy $J_1(k_{2l}a) \neq 0$, $J_1(k_{2l}a) \times J_2(k_{2l}a) > 0$ and $D_{21} = 0$. According to Eq. (7), $D_{21} = 0$ means that the normal projection of stress tensor is zero at the interface of the defect. Imposed by the elastic boundary condition, $P_1$ should be zero anywhere as $P_1$ is constant in the region (1). Thus, the incident wave will be totally blocked, which indicates total reflection. As expected, one can see from Fig. 2(b) that the pressure field $P_1$ disappears inside region (1) and the normal projection of stress tensor is zero at the interface indeed. Fig. 2(c) and 2(d) show control simulations in which ZIM is removed. It can be seen that in waveguides without ZIM most of the incidence waves are blocked by the large defect while more incidence waves go though as the defect diminishes, which is in accordance with common sense. However, comparing all the panels in Fig. 2, one can find an interesting fact that, under the proper conditions, ZIM will allow incidence waves totally go though the large defect while totally block the incidence waves even the defect is small.

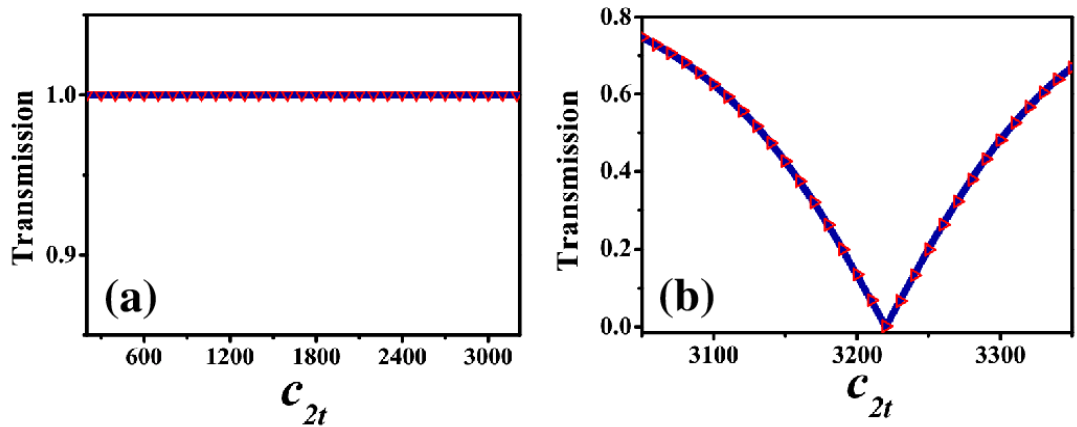

Figure 3. (Color online) Transmission coefficients as a function of the transverse wave speed of the steel defect for the ZIM waveguide structure realizing total transmission (a) and total reflection (b). The solid line represents the numerically calculated result, while the triangles represent the theoretical result.

Fig. 3(a) and 3(b) show the transmission coefficients as a function of the transverse wave speed of the steel defect for the ZIM waveguide structures realizing total transmission and total reflection, respectively. The solid line represents the numerically calculated result, while the triangles represent the theoretical result obtained from Eq. (8). The simulated result agrees with the theoretic result excellently, which further confirms our analysis. Here, we have assumed the transverse wave speed of the defect can be adjusted freely and we note that in Fig. 3(b) the speed at which $T = 0$ equal the transverse wave speed of steel exactly.

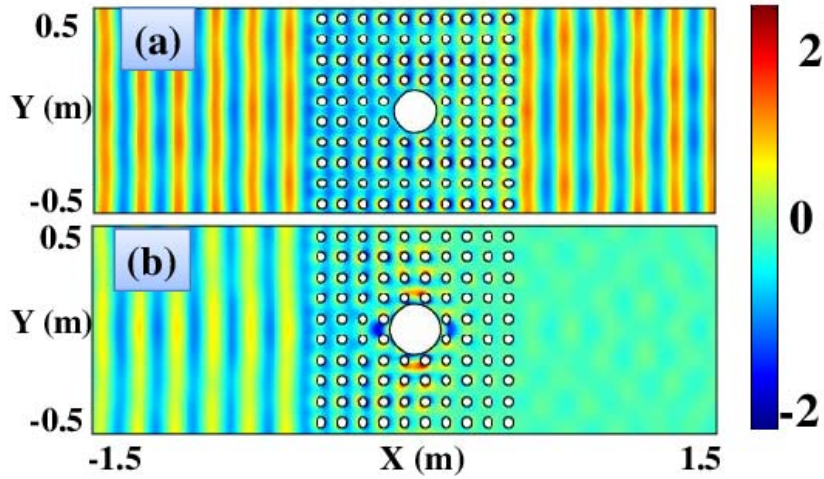

Figure 4. (Color online) The numerically simulated pressure field distributions for the waveguide structures with the 2D ACs system with the silicone rubber defect to achieve total transmission (a), the PMMA defect to achieve total reflection (b).

In our previous work[17], we have shown that a 2D ACs system consisting of rubber cylinders in water can be mapped to an acoustic material with effectively zero density and zero reciprocal modulus. Here, we use such 2D ACs system to demonstrate the intriguing transmission properties of the ZIM waveguide structure. For the convenience of simulation, we have replaced the steel defect with the silicone rubber defect and the PMMA defect to achieve total transmission and total reflection, respectively. Fig. 4 shows the numerically simulated results, and it can be seen that the ACs system is a good candidate to achieve the ZIM waveguide structure experimentally.

## IV. SUMMARY

In conclusion, we have demonstrated how to achieve total reflection and total transmission of acoustic waves by embedding a general solid defect in the ZIM waveguide. Theoretical analysis shows that adjusting transverse wave speed can only be applied to achieve total reflection instead of total transmission. Numerical simulations confirm our theory. We would like to point out that our result can be extended to the ZIM waveguide embedded with multiple defects. Our work provides more choices for manipulating acoustic wave propagation in ZIM with embedded general solid defects.

**Acknowledgement:**

The author acknowledges helpful discussions with Prof. Zhengyou Liu.